\documentclass[a4paper,12pt]{article}
\usepackage{amsfonts,amsthm,amsmath,amssymb,graphicx,hyperref,color,youngtab}
\newcommand{\be}{\begin{equation}}
\newcommand{\bea}{\begin{eqnarray}}

\newcommand{\ee}{\end{equation}}
\newcommand{\eea}{\end{eqnarray}}

\begin{document}

\makeatletter
\@addtoreset{equation}{section}
\makeatother
\renewcommand{\theequation}{\thesection.\arabic{equation}}

\rightline{}%WITS-CTP-108
\vspace{1.8truecm}

\vspace{15pt}

%%%%%%%%%%%%%%%%%

{\LARGE{  
\centerline{\bf Inelastic Magnon Scattering} 
}}  

\vskip.5cm 

\thispagestyle{empty}
    {\large \bf 
\centerline{Robert de Mello Koch\footnote{ {\tt robert@neo.phys.wits.ac.za}}, 
 and   Hendrik J.R. van Zyl\footnote{ {\tt hjrvanzyl@gmail.com}}}}

\vspace{.4cm}
\centerline{{\it National Institute for Theoretical Physics ,}}
\centerline{{\it School of Physics and Mandelstam Institute for Theoretical Physics,}}
\centerline{{\it University of Witwatersrand, Wits, 2050, } }
\centerline{{\it South Africa } }

\vspace{1.4truecm}

%%%%%%%%%%%%%%%%%
\thispagestyle{empty}

\centerline{\bf ABSTRACT}

\vskip.4cm 
We study the worldsheet $S$-matrix of a string attached to a D-brane in AdS$_5\times$S$^5$. 
The D-brane is either a giant graviton or a dual giant graviton. 
In the gauge theory, the operators we consider belong to the $su(2|3)$ sector of the theory.
Magnon excitations of open strings can exhibit both elastic (when magnons in the bulk of the string scatter) and inelastic (when
magnons at the endpoint of an open string participate) scattering.
Both of these $S$-matrices are determined (up to an overall phase) by the $su(2|2)^2$ global symmetry of the theory.
In this note we study the $S$-matrix for inelastic scattering.
We show that it exhibits poles correponding to boundstates of bulk and boundary magnons.
A crossing equation is derived for the overall phase.
It reproduces the crossing equation for maximal giant gravitons, in the appropriate limit.
Finally, scattering in the $su(2)$ sector is computed to two loops.
This two loop result, which determines the overall phase to two loops, will be useful when a unique solution to the 
crossing equation is to be selected. 

\setcounter{page}{0}
\setcounter{tocdepth}{2}

\newpage

\tableofcontents

\setcounter{footnote}{0}

\linespread{1.1}
\parskip 4pt

{}~
{}~

\section{Introduction}

't Hoofts original proposal that the large $N$ expansion of Yang-Mills theories are equivalent to a string theory\cite{'tHooft:1973jz}
is realized beautifully in the AdS/CFT correspondence\cite{Maldacena:1997re}.
Many concrete details of the duality can be confirmed with precision checks, thanks to integrability of planar 
${\cal N}=4$ super Yang-Mills theory\cite{Minahan:2002ve,Beisert:2010jr}:
the planar diagrams give rise to a two dimensional effective theory which can be matched, in exquisite detail, to the
worldsheet theory of a string.
This detailed matching is possible because integrability allows the exact $\lambda =g_{YM}^2N$ dependence of certain
quantities to be computed. 

There are many interesting string theory questions whose answer require the study of certain large $N$ but non-planar limits
of Yang-Mills theory.
One such example is the study of the open string excitations of giant graviton branes.
The physics of this problem requires summing many non-planar diagrams in the Yang-Mills theory, and so, corresponds
to non-perturbative string effects\cite{Balasubramanian:2001nh}.
Further, since the system is not in general integrable\cite{Koch:2015pga}, 
a detailed comparison akin to what was achieved in the planar limit seems impossible.
However, if one restricts to the $su(2|3)$ sector of the theory it turns out that the exact $S$ matrix describing the scattering 
of worldsheet excitations can still be determined up to an overall phase, by making use of the global $su(2|2)^2$ symmetry 
enjoyed by this sector\cite{Beisert:2005tm,Beisert:2006qh}.
The scattering of magnon excitations of open strings is inelastic\cite{Koch:2015pga}, which is a strong hint that 
the system is not integrable.

The fact that some quantities can be computed exactly, even without integrability, is extremely interesting and deserves
to be explored in detail.
The first goal of this study is to explore the structure of the $S$-matrix for inelastic magnon scattering and verify that
it has the stucture we expect.
Specifically, the analyticity and unitarity of the $S$-matrix implies a correspondence between singularities of the $S$-matrix 
and on-shell intermediate states. 
This is the subject of section \ref{bs}.
We find a pole corresponding to binding a bulk and a boundary magnon.
The structure of boundstates that we uncover smoothly interpolates between the bound state structure of bulk 
magnons\cite{Dorey:2006dq,Chen:2006gq} (for small giant gravitons when $r\approx 1$) and the bound state
structure obtained for maximal giants\cite{Hofman:2007xp} (when $r\approx 0$).
The boundstate is a BPS state in the double box representation of $su(2|2)^2$.
The second goal of this work is to study the overall phase of the $S$-matrix.
This phase is constrained by a crossing symmetry equation\cite{Janik:2006dc,Beisert:2006ez}.
Using insights following from similar studies of the same question in the planar limit \cite{Beisert:2006qh,Hofman:2007xp}, 
we write down an equation obeyed by this phase, by considering the scattering of a magnon with a singlet state.
Although we have not managed to solve this equation, we have checked that it reduces the crossing 
equation\cite{Hofman:2007xp} obtained for maximal giant gravitons in an appropriate limit.
Finally, we study scattering in the $su(2)$ sector, perturbatively to two loops, in the super Yang-Mills theory.
These results can be used to single out a unique solution to the crossing equation.

\section{Bound State Spectrum}\label{bs}

The scattering problem is most conveniently described using complex spectral parameters $x^\pm$.
In terms of these parameters, the charge, energy and momentum of a magnon can be written as follows\cite{Koch:2015pga}
\bea
n&=&{g\over i}\left( x^++{1\over x^+}-r x^--{r\over x^-}\right)\label{charge}
\eea
\bea
E&=&{g\over i}\left( x^+-{1\over x^+}-r x^-+{r\over x^-}\right)\label{energy}\qquad\qquad
e^{ip}={x^+\over x^-}
\eea
where $r=1$ for a bulk magnon, $0\le r <1$ for a boundary magnon attached to a giant graviton and
$r>1$ for a boundary magnon attached to a dual giant graviton.
Using the above relations we can determine the energy of a magnon in terms of its charge and momentum as
\bea
E=\sqrt{n^2+4 g^2 (1+r^2)-8g^2 r \cos (p)}\label{genenergy}
\eea
The condition that $x^+-r x^-$ and ${1\over x^+}-{r\over x^-}$ are pure imaginary will ensure real charges, energies,
and momenta.
To look for boundstates we will analyticaly continue the spectral parameters by relaxing this condition, allowing the 
energy and momenta to be complex.  
We will however maintain (\ref{charge}): this is the condition to have a short (atypical) representation of $su(2|2)^2$.
It is only for these representations that the tensor product of two representations is irreducible and hence that the 
$su(2|2)^2$ symmetry is sufficient to fix the $S$-matrix up to an overall phase.
The inverse relation is
\bea
x^\pm = {ie^{\pm i{p\over 2}}(E+n)\over 2g (e^{i{p\over 2}}-re^{-i{p\over 2}})}\label{spectralparam}
\eea
The scattering of a bulk and a boundary magnon is inelastic, as we now explain.
We use a subscript 1 to denote the bulk magnon before scattering and a subscript 2 to denote a boundary magnon
before scattering.
We used primed subscripts for the magnons after scattering.
The momenta, energies and charges after scattering are determined by solving
\bea
   E_1+E_2=E_1'+E_2'\qquad    p_1+p_2=p_1'+p_2'\qquad   n_1+n_2=n_1'+n_2'
\eea
These equations can be reduced to the solution of a cubic equation that has a single real root, but the details are not very
illuminating.
Close to $r=0$ and $r=1$ we can do better though: at $r=1-\epsilon$ we have
\bea
x_{1}^{\prime\pm}=x_{2}^\pm+\delta x_{1}^{\prime\pm}\qquad
x_{2}^{\prime\pm}=x_{1}^\pm+\delta x_{2}^{\prime\pm}\label{nearris1}
\eea
while at $r=\epsilon$ we have
\bea
x_{1}^{\prime\pm}=-x_1^\mp+\delta x_{1}^{\prime\pm}\qquad
x_{2}^{\prime +}=x_2^++\delta x_{2}^{\prime +}\qquad
x_{2}^{\prime -}=\left({x_1^-\over x_1^+}\right)^2 x_2^-+\delta x_{2}^{\prime -}\label{nearris0}
\eea
Working to order $\epsilon$ we find a set of linear equations whose explicit solution shows that
$\delta x_{1}^{\prime\pm}\sim O(\epsilon )$, $\delta x_{2}^{\prime\pm}\sim O(\epsilon)$ for both cases.
The linear equations we use arise by assuming we scatter elementary magnons (so that $n_1=n_2=n_{1}'=n_2'=1$),
as well as energy and momentum conservation.
The fact that $\delta x_{1}^{\prime\pm}$, $\delta x_{2}^{\prime\pm}$ are non-zero is a clear indication that the scattering
is not elastic.

We now focus attention on the scattering of magnons that belong to the $su(2)$ sector of the theory.
This is still perfectly general since the global symmetry of the theory then determines scattering in any 
other sector\cite{Beisert:2005tm}.
In this case, up to an undetermined overall phase, the $S$-matrix is given by\footnote{Here, following \cite{Koch:2015pga}, we 
use the notation $R$ to denote the $S$-matrix for the scattering of a bulk and a boundary magnon. 
We reserve $S$ for the $S$-matrix of bulk magnon scattering, which is an elastic process.}
\bea
R|\phi^1_1\phi^1_2\rangle = A^R_{12} |\phi^1_{1'}\phi^1_{2'}\rangle
\eea
where
\bea
A^R_{12}=R^0_{12}
\frac{\eta_1 \eta_2 x_1^{\prime +}x_1^+ (x_1^--x_2^+)\left((x_2^+-rx_2^-)(rx_2^{\prime +}-x_2^{\prime -})x_2^+
+(x_2^--rx_2^+)(x_2^{\prime +}-rx_2^{\prime -})x_2^{\prime +}\right)}{\eta_1' \eta_2' x_2^{\prime +}x_2^+
(x_1^--x_1^+)(x_1^+-x_1^{\prime +})(x_1^+(rx_2^+-x_2^-)+x_2^-(rx_2^--x_2^+))}\cr
\label{Smatrix}
\eea
This reduces to the correct bulk\cite{Beisert:2005tm} and reflection\cite{Hofman:2007xp} matrices when we set $r=1$ and $r=0$ respectively\footnote{To make this comparison we found \cite{Correa:2009dm} very useful.}.
The statement that the $S$-matrix is unitary is the statement
\bea
A_{12}^R A_{1'2'}^R =1
\eea
which we have verified holds for any $r$, as it should.

We will now look for singularities in the $S$-matrix. 
The presence of simple poles indicate on shell intermeddiate states.
Investigations of the singularities of the elastic magnon $S$-matrix has uncovered a wealth of BPS 
boundstates \cite{Dorey:2006dq,Chen:2006gq,Dorey:2007xn}.
We will argue below that we find an equally rich spectrum of BPS boundstates that naturally interpolate between the
boundstates of bulk magnons\cite{Dorey:2006dq} and the boundstates a bulk magnon 
and a boundary magnon associated to a maximal giant graviton\cite{Hofman:2007xp}. 
Inspection of (\ref{Smatrix}) suggests possible poles when $x_2^{\prime +}=0$ or when $x_2^+=0$.
Since the charges $n_k$ are positive integers and since we want to keep $Re(E_k)>0$, its clear from
(\ref{spectralparam}) that  these poles can't be realized.
The factor in the denominator $(x_1^--x_1^+)(x_1^+-x_1^{\prime +})$ can also give rise to a pole.
Analyzing this factor near $r=1$ we find a pole at
\bea
   x_1^+=x_2^++2 (1-r)x_2^+{x_2^+ x_2^--1\over (x_2^+-{1\over x_2^+})(x_2^+-x_2^-)}
+O\left( (1-r)^2\right)
\eea
This is precisely canceled, by a zero coming from the factor 
$(x_2^+-rx_2^-)(rx_2^{\prime +}-x_2^{\prime -})x_2^+
+(x_2^--rx_2^+)(x_2^{\prime +}-rx_2^{\prime -})x_2^{\prime +}$
in the numerator.
Near $r=0$ the factor $(x_1^--x_1^+)(x_1^+-x_1^{\prime +})$ in the denominator leads to a pole at
\bea
x_1^-=-x_1^++r{x_1^+\over x_2^-}
{(x_1^+-{1\over x_1^+})(x_2^--x_2^+)(x_2^-+x_2^+)\over x_2^- (x_2^+-{1\over x_2^+})}
+O\left(r^2\right)
\eea
This is again canceled, by a zero coming from the factor 
$(x_2^+-rx_2^-)(rx_2^{\prime +}-x_2^{\prime -})x_2^+
+(x_2^--rx_2^+)(x_2^{\prime +}-rx_2^{\prime -})x_2^{\prime +}$
in the numerator.
Thus, in the end we find that a single pole arises when $x_1^+(rx_2^+-x_2^-)+x_2^-(rx_2^--x_2^+)=0$, which implies that
\bea
x_1^+=x_2^-{x_2^+-rx_2^-\over rx_2^+-x_2^-}\label{bscondition}
\eea
To interpret this pole recall that singularities of the $S$-matrix correspond to spacetime diagrams where each particle is 
on-shell\cite{Iagolnitzer:1978qv}. 
Particle worldlines meet at vertices which conserve charge, energy and momentum.
We want to consider a cubic vertex corresponding to the creation of a boundstate from a boundary and a bulk magnon.
Using $b'$ to denote the boundstate of a boundary and a bulk magnon, the conservation of charge, energy and momentum
implies that
\bea
 E(x_1^\pm)+E(x_2^\pm)&=&E(x_{b'}^\pm)\cr
 p(x_1^\pm)+p(x_2^\pm)&=&p(x_{b'}^\pm)\cr
 n(x_1^\pm)+n(x_2^\pm)&=&n(x_{b'}^\pm)
\eea
Since this is three equations we can completely determine the spectral parameters of the boundstate using two of the
above equations.
The third equation then implies a relation between the magnon and boundary magnon spectral relations that is obeyed
by (\ref{bscondition}).
It is straight forward to apply the rules described in \cite{Dorey:2007xn} and verify that this pole signals a normalizable wave function,
for a boundstate with charge $n=2$ and energy given by (\ref{genenergy}) evaluated at this $n$.
Further, when $r=1$ (\ref{bscondition}) is the pole identified in \cite{Dorey:2006dq} as a signal of a bound state of 
bulk magnons and when $r=0$ (\ref{bscondition}) is the pole identified in \cite{Hofman:2007xp} as the signal of a
bound state of a bulk and a boundary magnon, for the case of open string attached to a maximal giant graviton.

We can now continue and consider the scattering of a bulk magnon with this boundate, producing a new boundstate
with charge $n=3$. 
Indeed, following the rules of \cite{Dorey:2007xn}, we find a family of boundstate with one boundary magnon 
bound to $n-1$ bulk magnons.
This boundstate has a charge of $n$ and energy given by (\ref{genenergy}).
By varying $r$ smoothly from $r=0$ to $r=1$, this structure of boundstates nicely interpolates from the known structure of
boundstates when bulk magnons bind to the boundary magnon of a maximal giant graviton\cite{Hofman:2007xp} and the known
structure for bulk magnons boundstates\cite{Dorey:2006dq}.

\section{Crossing Equation}\label{cross}

To derive the crossing equation, we will follow the derivation given in \cite{Beisert:2006qh}.
The same method has been applied to determine the crossing equation for maximal giant gravitons in \cite{Hofman:2007xp}.
The idea is to consider the scattering from the singlet state
\bea
   |I_{\bar{p},p}\rangle = f_p\left(|\phi^1_{\bar{p}}\phi^2_p\rangle - |\phi^2_{\bar{p}}\phi^1_p\rangle\right)+
                                          |\psi^1_{\bar{p}}\psi^2_p\rangle - |\psi^2_{\bar{p}}\psi^1_p\rangle
\eea
The phase factor $f_p$ is determined by requiring that the singlet state is annihilated by all of the $su(2|2)$ elements, as 
explained in \cite{Beisert:2006qh}.
To obtain the crossing equation, we consider the scattering of this impurity from the right boundary magnon.
The only difference we find as compared to \cite{Hofman:2007xp}, is that the scattering is inelsatic.
Since the scattering is inelastic, we find
\bea
  R(\bar{p},q')S(\bar{p},p')R(p,q)|I_{\bar{p},p},q\rangle = \phi_r |I_{p',\bar{p}'},q\rangle
\eea
where $r$ is a phase. 
If we now scatter the singlet from the left boundary magnon, explicit computation show that we pick up the same phase, so
that we return to the original state apart from the phase $\phi_r^2$.
The crossing equation is now obtained by requiring that $\phi_r^2$ is one, i.e. that $\phi_r=\pm 1$.
To choose the correct sign, we compare to the $r=0$ limit (studied in \cite{Hofman:2007xp}) and conclude 
that we should impose $\phi_r=1$.
Since this crossing equation involves both the scattering of bulk with bulk magnons and the scattering of boundary with bulk
magnons, it relates the overall phase factor of $S$ (which has been determined) to the overall phase factor of $R$ (what we want
to determine).
After a tedious computation we find the following result
\bea
   {\left(
f_p C^R (p, q)G^S (\bar{p},p')-2L^R (p, q)K^S(\bar{p}, p'))
(H^R(\bar{p}, q')K^R(\bar{p}, q')-G^R(p, q')L^R(\bar{p}, q')\right)\over 2L^R(\bar{p},q')}=1
\cr
\label{crossres}
\eea
where the matrix elements of $R$ (denoted with superscript $R$) are derived in \cite{Koch:2015pga} and the matrix elements
of $S$ (denoted with superscript $S$) are derived in \cite{Beisert:2006qh}.
We will refer to the function on the LHS of (\ref{crossres}) as the crossing function.
Scattering with different boundary magnon states determines crossing functions that have a different expression in terms of the
matrix elements of $R$ and $S$, but lead to the same crossing equation.
We have not written this crossing equation in terms of the spectral parameters of the initial and final magnons as these
expressions are rather long.
The phase we have discussed above arises from an $su(2|2)$ factor. 
The theory actually enjoys $su(2|2)^2$ symmetry, so the full reflection factor is the square of the phase factor we discussed
above.

As a first check of these results, we note that when $r=0$ they reproduce the crossing equation quoted in
equation (41) of  \cite{Chen:2007ec}.
In addition to this, a numerical study of the crossing equation reveals an appealing symmetry.
Recall that we denote the boundary magnon momentum by $q$ and the bulk magnon momentum by $p$.
The crossing equation obtained from scattering off the right boundary with momenta $ p + q \to  p' + q'$ is identical to 
crossing equation obtained the left scattering with the same momenta $ q + p \to q' + p'$.  
If one considers right scattering with momenta $p' + q' \to p + q$ or left scattering with momenta $q' + p' \to q + p$ the 
crossing equations are again identical: for all four situations we obtain the same crossing equation.
The appearance of this symmetry is important for the consistency of our derivation of the crossing equation since it
ensures that scattering the singlet from both boundaries as described above, does indeed return us to our initial state.
It is satisfying that this derivation of the crossing equation works even though we have inelastic scattering.

\section{$su(2)$ Scattering to Two Loops}

We will now consider the scattering of a bulk and a boundary magnon, at two loops, in the super Yang-Mills theory.
This will allow us to determine the overall phase of $R$ to two loops, which will be useful data when a unique solution
to the crossing equation is to be singled out.
The operators in the Yang-Mills theory dual to giant gravitons are given by Schur 
polynomials \cite{Corley:2001zk,Corley:2002mj,Berenstein:2004kk}.
Giant gravitons with open string excitations are dual to the restricted Schur polynomials,
constructed in \cite{Balasubramanian:2004nb,de Mello Koch:2007uu}.
The action of the dilatation operator on the restricted Schur polynomials has been constructed in 
\cite{de Mello Koch:2007uv,Bekker:2007ea,Koch:2015pga}.
In what follows below we use this action at two loops to define the Hamiltonian for a Schr\"odinger equation based description
of the magnon scattering.
The Bethe ansatz for the wave function is given by\cite{Staudacher:2004tk}
\bea
    \psi (l_1,l_2)=e^{ip_1 l_1+ip_2l_2}+A e^{ip'_1 l_1 +ip_2' l_2}+g^2 \phi (l_1)\delta_{|l_1-l_2|,1}
\eea
where if $|l_1-l_2|>2$ the wave function must obey the Schr\"odinger equation
\bea
E\Psi (l_1,l_2)&=&
g^2(3+r^2)\Psi(l_1,l_2)
-g^2 r \left(\Psi(l_1-1,l_2)+\Psi(l_1+1,l_2)\right)\cr
&&-g^2\left(\Psi(l_1,l_2-1)+\Psi(l_1,l_2+1)\right)
-g^4 \left(\left(1+r^2 \right)^2+4\right)\Psi (l_1,l_2)\cr
&&+2g^4\left(1+ r^2 \right)r\left(\Psi(l_1-1,l_2)+\Psi(l_1+1,l_2)\right)\cr
&&+4g^4\left(\Psi(l_1,l_2-1)+\Psi(l_1,l_2+1)\right)\cr
&&-g^4 r^2\left(\Psi(l_1-2,l_2)+2\Psi(l_1,l_2)+\Psi(l_1+2,l_2)\right)\cr
&&-g^4 \left(\Psi(l_1,l_2-2)+2\Psi(l_1,l_2)+\Psi(l_1,l_2+2)\right),\label{tdepse}
\eea
$l_1$ is the position of the boundary magnon, with momentum $p_1$.
$l_2$ is the position of the bulk magnon, with momentum $p_2$.
If $l_2=l_1+2$, the Schr\"odinger equation becomes
\bea
E\Psi (l_1,l_1+2)&=&
g^2(3+r^2)\Psi(l_1,l_1+2)
-g^2r \left(\Psi(l_1-1,l_1+2)+\Psi(l_1+1,l_1+2)\right)\cr
&&-g^2\left(\Psi(l_1,l_1+1)+\Psi(l_1,l_1+3)\right)
-g^4 \left(\left(1+r^2 \right)^2+4\right)\Psi (l_1,l_1+2)\cr
&&+2g^4\left(1+r^2 \right)r\left(\Psi(l_1-1,l_1+2)+\Psi(l_1+1,l_1+2)\right)\cr
&&+4g^4\left(\Psi(l_1,l_1+1)+\Psi(l_1,l_1+3)\right)
-g^4 \left(4\Psi(l_1,l_1+2)+\Psi(l_1,l_1+4)\right)\cr
&&-g^4 r^2 \left(\Psi(l_1-2,l_1+2)+2\Psi(l_1,l_1+2)\right),\label{frstspsc}
\eea
and if $l_2=l_1+1$, the Schr\"odinger equation becomes
\bea
E&&\!\!\!\!\!\!\!\!\!\Psi (l_1,l_1+1)=
g^2(1+r^2)\Psi(l_1,l_1+1)
-g^2r\Psi(l_1-1,l_1+1)-g^2\Psi(l_1,l_1+2)\cr
&&-g^4 \left(r^4+1\right)\Psi (l_1,l_1+1)
+2g^4 \left(1+r^2\right)r\Psi(l_1-1,l_1+1)+4g^4\Psi(l_1,l_1+2)\cr
&&-g^4 r^2\left(\Psi(l_1-2,l_1+1)+\Psi(l_1,l_1+1)\right)
-g^4 \left(\Psi(l_1,l_1+1)+\Psi(l_1,l_1+3)\right)\cr
&&-g^4r\left(\Psi (l_1-1,l_1)+\Psi (l_1+1,l_1+2)\right),\label{scndspsc}
\eea
From (\ref{tdepse}) we learn that
\bea
E=g^2(3+r^2)-g^2 r (e^{-ip_1}+e^{ip_1})-g^2(e^{-ip_2}+e^{ip_2})
-g^4 (1+r^2)^2-4g^4\cr
+2g^4 (1+r^2)r (e^{-ip_1}+e^{ip_1})+4g^4 (e^{-ip_2}+e^{ip_2})
-g^4 (e^{-2ip_2}+2+e^{2ip_2})\cr
-g^4 r^2 (e^{-2ip_1}+2+e^{2ip_1})
\eea
From (\ref{frstspsc}) we find
\bea
r \phi (l_1+1)+\phi (l_1)=r^2 \psi (l_1+2, l_1+2)+\psi(l_1,l_1)-2\psi (l_1,l_1+2)\label{Aeqn}
\eea
Finally, from (\ref{scndspsc}) we find
\bea
g^4 (2-r(e^{-ip_1}+e^{ip_1})-e^{ip_2}-e^{-ip_2})\phi (l_1)\cr
=-g^2 (g^2 r \psi(l_1-1, l_1) - g^2 \psi(l_1,l_1 -1) - \psi(l_1, l_1) + 4 g^2 \psi(l_1, l_1)\cr
 + 2 \psi (l_1, l_1+1) - 5 g^2 \psi(l_1,  l_1+1) - 3 g^2 r^2 \psi (l_1, l_1+1) - r \psi (l_1+1,l_1+1)\cr
 +    2 g^2 r \psi (l_1+1,l_1+1)+2 g^2 r^3 \psi (l_1+1,l_1+1) +
    g^2 r \psi(l_1+1, l_1+2) - g^2 r^2 \psi (l_1+2,l_1+1))
\label{peqn}
\eea
Starting from (\ref{peqn}) we are able to solve for $\phi (l_1)$. 
We find that $\phi (l_1)$ is independent of $l_1$ which is intuitively appealing.
Inserting the solution for $\phi(l_1)$ into (\ref{Aeqn}), we are able to solve for $A$. 
The result is

\vfill\eject

\begin{eqnarray}
A =  -\frac{1 - 2 e^{i p_2} + e^{i (p_1 + p_2)}r}{1 - 2 e^{i p_2'} + e^{i(p_1' + p_2')}r} - \frac{g^2(e^{i p_1} - e^{i p_1'})}{(1 + r e^{i(p1 + p2)})(1 - 2e^{i p_2'} + r e^{i (p_1' + p_2')}   )^2} \times \cr
\big[2\left(r e^{i (p_1+p_2)}-2 e^{i p_2}+1\right) \left(r e^{i (2 p_1+p_2)}-2 e^{i (p_1+p_2)}
+e^{i (p_1+2 p_2)}+e^{i p_1}+e^{i p_2} r\right) e^{i \left(-p_1'-p_1+p_2'\right)}\cr
+e^{i \left(-p_1'-p_1\right)} \left(-e^{i p_2} \left(e^{i p_1} \left(r^2-9\right)+4 e^{2 i p_1} r+2 r\right)
-e^{2 i p_2} \left(4 e^{3 i p_1} r^2+e^{2 i p_1} \left(r^2-7\right) r+6 e^{i p_1}-2 r\right)\right.\cr
+e^{i (p_1+3 p_2)} \left(2+r \left(-4 e^{3 i p_1} r^2+e^{2 i p_1} \left(r^2+7\right) r-2
   e^{i p_1} \left(r^2+2\right)+2 r\right)\right)\cr
\left.+r e^{2 i (p_1+2 p_2)} \left(2+e^{i p_1} r \left(-2+e^{i p_1} r\right)\right)-4 e^{i
   p_1}+r\right)\cr
+\left(1+r e^{i (p_1+p_2)}\right) \left(r e^{i (p_1+p_2)}-2 e^{i p_2}+1\right) \left(r^2 e^{i
   (p_1+p_2)}+e^{i (-p_1-p_2)}\right)\big]\cr
\label{Aresult}
\end{eqnarray}

Recall that total $R$-matrix has a contribution from each of the $su(2|2)$ factors, so that
\bea
  R_{\rm R}(x_1,x_2,x_1',x_2')= R_{su(2|2)}(x_1,x_2,x_1',x_2')\otimes  R_{su(2|2)}(x_1,x_2,x_1',x_2')
\eea
Consequently, setting $A$ in (\ref{Aresult}) to be equal to $(A_{12}^R)^2$ with $A_{12}^R$ given in (\ref{Smatrix}),
we determine the overall phase to two loops.

{\vskip 0.5cm}

\noindent
{\it Acknowledgements:}
This work is based upon research supported by the South African Research Chairs
Initiative of the Department of Science and Technology and National Research Foundation.
Any opinion, findings and conclusions or recommendations expressed in this material
are those of the authors and therefore the NRF and DST do not accept any liability
with regard thereto.

\end{document}